\documentclass[draft]{agujournal2019}
\usepackage{url} 
\usepackage{lineno}
\usepackage[inline]{trackchanges} 
\usepackage{soul}
\usepackage{multirow}
\usepackage{booktabs}

\draftfalse

\journalname{Space Weather}

\begin{document}

%
%


\title{Predicting CME Arrivals with Heliospheric Imagers from L5: A Data Assimilation Approach}

%
%




\authors{T. Amerstorfer\affil{1}, 
            J. Le Louëdec\affil{1},
            D. Barnes\affil{2},
            M. Bauer\affil{1,3}, 
            J.A. Davies\affil{2}, 
            S. Majumdar\affil{1}, 
            E. Weiler\affil{1,3}, 
            C. Möstl\affil{1}}

\affiliation{1}{Austrian Space Weather Office, GeoSphere Austria, Reininghausstraße 3, 8020 Graz, Austria}
\affiliation{2}{RAL Space, STFC Rutherford Appleton Laboratory, Didcot, UK}
\affiliation{3}{Institute of Physics, University of Graz, Universit\"atsplatz 5, Graz, 8010, Austria}




\correspondingauthor{Tanja Amerstorfer}{tanja.amerstorfer@geosphere.at}



\begin{keypoints}
\item Incorporating HI observations up to at least $35^\circ$ elongation improves CME arrival predictions over coronagraph-based methods.
\item Reducing the prediction lead time from 100 h to 50 h corresponds to an improvement in forecast accuracy of roughly $45~\%$.
\item Access to high-quality real-time  HI data from future missions like Vigil at L5 will greatly enhance our ability to predict CMEs.
\end{keypoints}

%
%

%
%


\begin{abstract}
The Solar TErrestrial RElations Observatory (STEREO) mission has laid a foundation for advancing real-time space weather forecasting by enabling the evaluation of heliospheric imager (HI) data for predicting coronal mass ejection (CME) arrivals at Earth. This study employs the ELEvoHI model to assess how incorporating STEREO/HI data from the Lagrange 5 (L5) perspective can enhance prediction accuracy for CME arrival times and speeds. Our investigation, preparing for the upcoming ESA Vigil mission, explores whether the progressive incorporation of HI data in real-time enhances forecasting accuracy. 
The role of human tracking variability is evaluated by comparing predictions based on observations by three different scientists, highlighting the influence of manual biases on forecasting outcomes. Furthermore, the study examines the efficacy of deriving CME propagation directions using HI-specific methods versus coronagraph-based techniques, emphasising the trade-offs in prediction accuracy. Our results demonstrate the potential of HI data to significantly improve operational space weather forecasting when integrated with other observational platforms, especially when HI data from beyond 35$^\circ$ elongation are used. These findings pave the way for optimising real-time prediction methodologies, providing valuable groundwork for the forthcoming Vigil mission and enhancing preparedness for CME-driven space weather events.
\end{abstract}

\section*{Plain Language Summary}

Large eruptions from the Sun, called coronal mass ejections (CMEs), can send billions of tons of charged particles into space. If these eruptions head towards Earth, they can disrupt satellites, power grids, and communication systems. To better predict if and when these solar storms will arrive, numerous methods have been used to model the evolution of CMEs through space.
In this study, we test ways to incorporate real-time heliospheric imager data into the ELEvoHI model, which uses images from space telescopes that track CMEs as they travel away from the Sun. We assess whether assimilating more HI data during a CME's propagation towards Earth improves arrival time predictions.
We also examine how prediction accuracy is influenced by different people performing the CME tracking and compare results using various estimates of the initial CME parameters needed as input to ELEvoHI.
Our findings suggest that using images of CMEs until they reach at least 35$^\circ$ from the Sun improves predictions compared to relying only on earlier images, and that results can vary depending on the analyst.
This research is important for future space weather missions, especially Vigil, an ESA satellite under development that will monitor Earth-directed CMEs from the L5 point.

%
%

\section{Introduction}

Coronal mass ejections (CMEs) are the most significant drivers of space weather, capable of causing severe geomagnetic storms when they interact with Earth's magnetosphere \cite{schw05}. Accurate prediction of the arrival times and speeds of CMEs at Earth is crucial for mitigating their potential impacts on technological systems and human activities in space. Over the past decades, substantial progress has been made in forecasting CME arrival times using various observational data and modelling approaches. However, the accuracy of today's mean absolute arrival time error still lies around 13.2 hours with a bias of $-2.5$ hours as derived by \citeA{kay24} from 2617 real-time predictions of 474 events. 

Traditionally, coronagraph data, which provide images of CMEs as they erupt from the Sun up to a maximum of 30 solar radii, have been the cornerstone of CME forecasting efforts. These data allow for the estimation of key CME parameters, such as speed, direction, and angular width, which are then ingested into operational models like the WSA-ENLIL+Cone model \cite{arg03,ods03} to predict CME arrivals at Earth. Although input derived from coronagraph data is useful, such operational models often struggle with high uncertainties related to the initial CME parameters, leading to errors in arrival time predictions \cite{sin22,ver23}. Especially when using coronagraph data from only one vantage point, it is challenging to derive these CME parameters without significant ambiguity. An enlightening study was done by \citeA{palkay24}, who compiled an extensive list of CMEs fitted with the graduated cylindrical shell model \cite<GCS;>{the11}. For more than 500 events from 24 catalogues, they derived the accuracy of the fitted parameters, the most important (for this study) being the longitude of the CME apex with an uncertainty of $8^\circ$ and the angular width with an uncertainty of $9.3^\circ$.

Despite these challenges, coronagraph data have been fundamental in operational space weather forecasting, as it provides essential estimates of initial CME parameters. However, integration of additional data sources, such as from heliospheric imagers (HI) or in situ measurements, could further enhance the forecasting accuracy. A lot of development is ongoing within the science community, but so far none of these methods are integrated as standard into operational CME forecasts.

HI data have emerged as a valuable complement to coronagraph observations. Four missions are currently equipped with HI cameras that track CMEs as they propagate through the interplanetary space, providing measurements of the elongation of features such as the CME front. While data from the Wide-Field Imager for Solar Probe \cite<WISPR;>{vou16} onboard Parker Solar Probe \cite<PSP;>{fox16} and from the Solar Orbiter Heliospheric Imager \cite<SolOHI;>{how20} onboard Solar Orbiter \cite<SolO;>{mue20} are only available for specific time ranges, the Heliospheric Imagers \cite<HI;>{eyl09} onboard the Solar TErrestrial RElations Observatory \cite<STEREO;>{kai08} are constantly monitoring a region around the ecliptic between $4$ and $88^\circ$ elongation from Sun-centre since end of 2006. The Polarimeter to UNify the Corona and Heliosphere \cite<PUNCH;>{def25}, launched in March 2025, now provides continuous $360^\circ$ observations of the outer corona and inner heliosphere with 4 spacecraft from low Earth orbit, further adding valuable new capabilities to heliospheric imaging.

Tools like the ELlipse Evolution model based on HI observations \cite<ELEvoHI;>{rol16, ame18} leverage these data to refine predictions by accounting for the evolution of CME kinematics over time.
The combination of heliospheric imager data with traditional coronagraph observations could significantly improve the accuracy of CME arrival forecasts, although challenges remain in integrating these different data sources effectively. One of these challenges is the absence of HI data in a sufficient quality in real-time \cite{tuc15,bau21}. 
Current real-time STEREO/HI data often suffer from significant gaps, low spatial and temporal resolution, and intermittent coverage, making them difficult to use reliably for continuous forecasting operations. These limitations mean that, while HI data can improve predictive accuracy in controlled studies analysing well-observed CMEs, its practical utility in day-to-day space weather forecasting is constrained by data quality and availability issues making it almost impossible to compare HI-based model accuracy to those of models used in real-time.
\citeA{lel25} recently applied machine learning techniques to enhance the quality and continuity of real-time HI data, demonstrating how such approaches can mitigate data gaps and noise. Their work represents an important step towards making HI observations operationally useful for real-time CME forecasting.

Some of the challenges pertaining to real-time HI data will be addressed with the launch of the Vigil spacecraft planned for 2031. Vigil will be continuously observing the space between the Sun and Earth from the L5 point of the Sun-Earth system and thus will provide a side-view onto Earth-directed CMEs. The mission is specifically designed to meet space weather forecasting needs by providing real-time HI data that have adequate temporal and spatial resolution and a quality comparable to today's STEREO/HI science data.

Due to the large data heritage provided by STEREO we can focus presently on developing approaches based on the incorporation of STEREO/HI science data into real-time applications. This study focuses on the use of high-quality HI science data, as if it had been available to predict CME arrivals at Earth before impact. In view of the upcoming Vigil mission, we only use data from times when STEREO was approximately located around the L5 point.
By applying the CME propagation model, ELEvoHI, we test if incorporating progressively more HI data---as it becomes available---to update our prediction, leads to an improvement of the model accuracy in terms of arrival time and arrival speed.
Additionally, we compare two different approaches to determine the CME propagation direction needed as input to ELEvoHI: a method solely relying on HI data from a single vantage point (L5) and another method applied to coronagraph data from two perspectives (L5 and L1).
As tracing a CME front through the HI field of view (FOV) is usually done by hand, we also test the influence of three scientists tracking the same features on the prediction results.
Finally, we discuss the most effective way to utilise HI data in the future to enhance our current prediction capabilities.

\section{Data}
\label{sec:data}

The wide-angle heliospheric imager data collected by STEREO \cite{kai08} offers a valuable data set to be used for assessing the future space weather forecasting capabilities of Vigil. At a heliocentric distance of slightly less than 1 au, STEREO-Ahead (STEREO-A) leads the Earth in its orbit, while STEREO-Behind (STEREO-B), at a heliocentric distance slightly larger than 1 au, lags. The angular separation of the two spacecraft changes by roughly 45$^\circ$ per year. STEREO was designed to be a twin-satellite space endeavour and allows for triangulation methods \cite{liu10a,liu10b} when the same event is observed from both vantage points. During solar conjunction in 2014, communication with STEREO-B was lost, such that only data from STEREO-A remained available thereafter. 

STEREO/HI data, part of the Sun Earth Connection Coronal and Heliospheric Investigation \cite<SECCHI;>{how02} suite, consists of two wide angle cameras centred on the ecliptic plane. HI1 observes elongations in the range $4^\circ$--$24^\circ$, whereas HI2 extends the coverage to $18^\circ$--$88^\circ$. HI data undergoes preprocessing to enhance CME visibility and minimise residual noise. We utilise a Python implementation of the \texttt{secchi\_prep.pro} routine, originally developed in IDL SolarSoft, to process the images to Level~1. This includes masking saturated CCD columns, correcting for smearing effects due to the shutter-less readout, applying flat-field and distortion corrections. Lastly, corrected pointing information available in SolarSoft is applied to the header data. To further enhance transient structures, we subtract the 5-day pixel-wise median background from each image ensuring accurate alignment so that the star field remains fixed across frames. After this alignment, running difference images are generated by subtracting the previous frame from the current one, highlighting dynamic features while suppressing static background elements. 

To extract the time-elongation profile of a CME, we construct Jmaps \cite{she99, dav09b} from these running difference images by extracting 2$^\circ$-wide strips centred on the Earth ecliptic from each processed HI image and vertically stacking them sequentially. Using this method, solar transients show up as bright streaks in the resulting image, which allows for the tracking of CMEs as they move along the ecliptic plane. All Python-routines to process and track the time-elongation profile of a CME in STEREO/HI data are freely available (see Section~\ref{sec:data_access}).

Observations from the Large Angle Spectroscopic COronagraph \cite<LASCO;>{bru95} C2 and C3 on-board the Solar and Heliospheric Observatory \cite<SOHO;>{dom95} and COR2 coronagraphs on-board STEREO-A/B are also used in this work to gain the propagation direction and the longitudinal extent of the CMEs (described in more detail in Section~\ref{sec:GCS}), which is simplified by 
associating the CMEs with their source regions. For this, various passbands of the Atmospheric Imaging Assembly \cite<AIA;>{lem12} on-board the Solar Dynamics Observatory \cite<SDO;>{pes12} and the Extreme Ultraviolet Imager \cite<EUVI;>{wue04}, also part of the SECCHI-suite on-board the twin STEREO spacecraft, are used.

\subsection{Set of CME events}

Associating a CME with its in situ heliospheric counterpart, the so-called interplanetary CME (ICME), can be challenging. Especially in the case of multiple events leaving the Sun within a short time range, the one-to-one relation to an in situ detection can be hard if not impossible. We use the term ICME as in \citeA{rou11}, where it is defined as the whole disturbance in the solar wind. The ICME sheath, starting with a shock (if present), is a region of enhanced speed and proton number density and gives prior notice to the following flux rope that is characterised by a larger total magnetic field strength and helical field lines twisted around an axial field \cite{bur81}. Since we tend to follow the leading edge of the CME in heliospheric images, i.e.\ a region appearing brighter in white-light due to its higher density, we are comparing our modelling results to the time of the arrival of the ICME sheath and not to the start time of the magnetic flux rope.

In order to test and validate our forecasting approach for potential future use with Vigil data, we compile a list of single, isolated Earth-directed CME events that were observed remotely from a vantage point around L5 (either from STEREO-A or B) and showed a clear in situ signature at L1. We select each of the events based on their presence in the HELCATS HICAT catalogue (the official event list for all CMEs observed by the HI instruments) and an associated in situ arrival at L1. For further information on HICAT see \citeA{har18}. Table \ref{tab:events} lists the event sample including the HELCATS ID, the longitude of the HI observer, the first appearance in HI1 and the arrival time and speed at L1.

\begin{table}[h!]
\caption{Events used for this study. The columns show the number of event, the HELCATS ID, the separation of STEREO from Earth in degrees, the first data point in HI in UT, the time of in situ arrival in UT, and the speed at in situ arrival in km~s$^{-1}$. For events with an asterisk, a GCS fit was possible. For events marked with a dagger, benchmark results from CCMC's CME Scoreboard are available. The superscripts at the in situ arrival times mark the source of the CME-ICME association, which is 1 for the HELCATS LINKCAT, 2 for HELIO4CAST, and 3 for CME Scoreboard.}
\label{tab:events}
\centering
\begin{tabular}{l c c c c c}
\hline\hline
 Event & HELCATS ID & HI s/c Longitude & $t_\textsubscript{HI1}$ & $t_\textsubscript{L1}$ &   $v_\textsubscript{L1}$\\
 \hline
CME01$^*$ & HCME\_B\_20100203\_01 & $-71$ & 2010-02-03 21:03 & 2010-02-07 18:04$^1$ & 349 \\ 
CME02 & HCME\_B\_20100616\_02 & $-70$ & 2010-06-16 23:34 & 2010-06-21 03:35$^1$ & 405 \\ 
CME03$^*$ & HCME\_A\_20200415\_01 & $-75$ & 2020-04-15 21:17 & 2020-04-20 01:34$^2$ & 346 \\ 
CME04 & HCME\_A\_20200623\_01 & $-70$ & 2020-06-23 02:31 & 2020-06-30 01:12$^2$ & 332 \\ 
CME05 & HCME\_A\_20200930\_01 & $-61$ & 2020-09-30 18:24 & 2020-10-05 06:52$^2$ & 353 \\ 
CME06$^*{^\dagger}$ & HCME\_A\_20201207\_01 & $-57$ & 2020-12-07 19:02 & 2020-12-10 02:10$^3$ & 481 \\ 
CME07$^*$ & HCME\_A\_20210211\_01 & $-56$ & 2021-02-11 01:42 & 2021-02-15 18:58$^2$ & 354 \\ 
CME08$^*{^\dagger}$ & HCME\_A\_20210220\_01 & $-56$ & 2021-02-20 19:47 & 2021-02-24 04:08$^2$ & 481 \\ 
CME09 & HCME\_A\_20210410\_01 & $-54$ & 2021-04-10 18:53 & 2021-04-15 03:28$^3$ & 383 \\ 
CME10${^\dagger}$ & HCME\_A\_20210422\_01 & $-53$ & 2021-04-22 09:33 & 2021-04-24 22:24$^3$ & 478 \\ 
CME11${^\dagger}$ & HCME\_A\_20210509\_01 & $-52$ & 2021-05-09 14:59 & 2021-05-12 05:48$^3$ & 445 \\ 
CME12$^*{^\dagger}$ & HCME\_A\_20210529\_01 & $-51$ & 2021-05-29 01:49 & 2021-06-02 12:20$^3$ & 308 \\ 
CME13$^*{^\dagger}$ & HCME\_A\_20210823\_01 & $-43$ & 2021-08-23 13:36 & 2021-08-27 00:25$^3$ & 406 \\ 
CME14$^*{^\dagger}$ & HCME\_A\_20210913\_01 & $-41$ & 2021-09-13 16:22 & 2021-09-17 01:29$^3$ & 340 \\ 
CME15$^*{^\dagger}$ & HCME\_A\_20211009\_01 & $-39$ & 2021-10-09 09:46 & 2021-10-12 01:46$^3$ & 441 \\ 
\hline
\end{tabular}
\end{table}

\section{Methods}

\subsection{Tracking approach in heliospheric imager data}

\begin{figure}[h!]
\includegraphics[width=\textwidth]{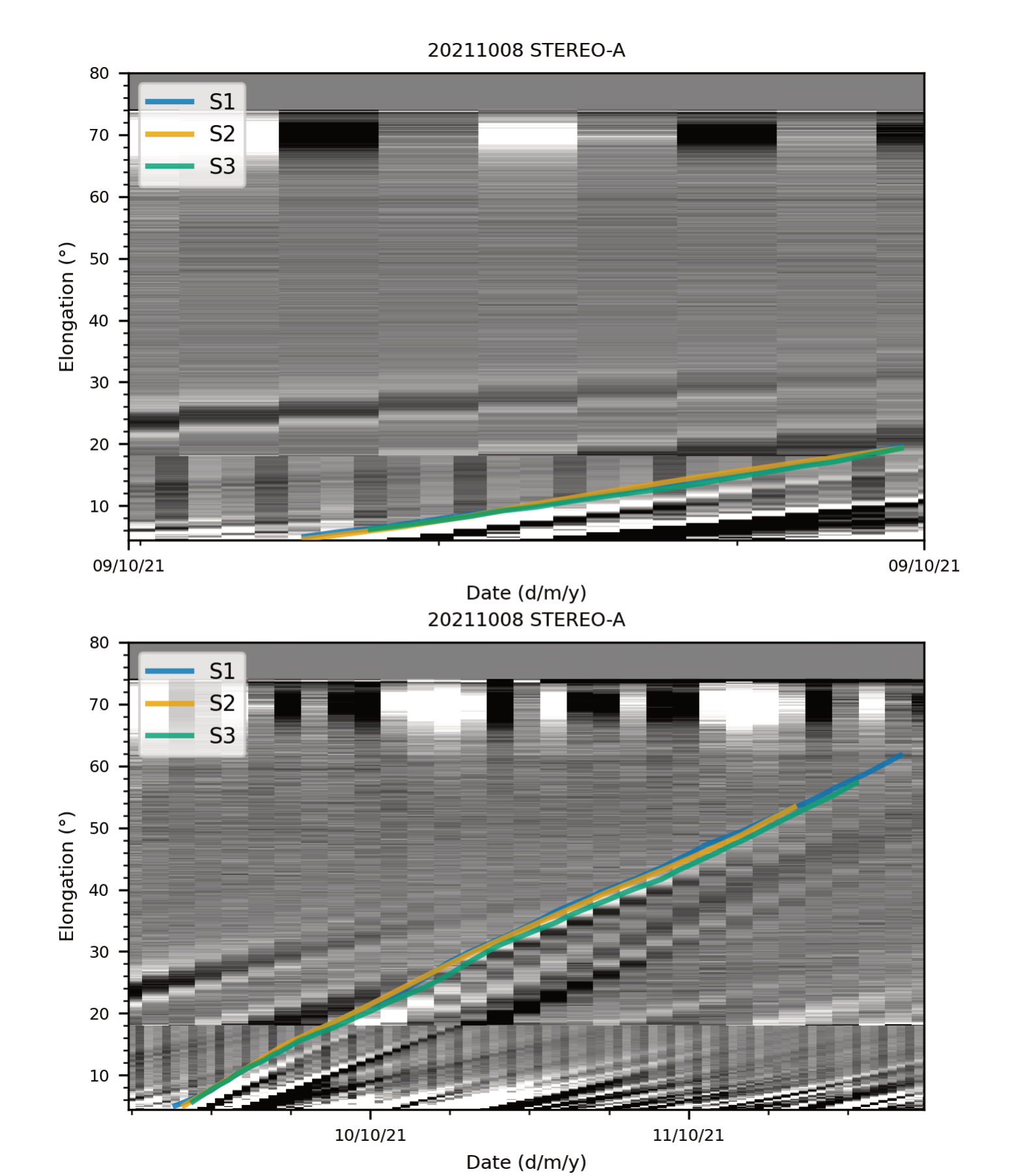}
\caption{Examples of time-elongation plots in the ecliptic plane showing the minimum (upper panel) and maximum (lower panel) amounts of available HI data for event CME15. The mean track obtained by each scientist is overlaid.}
\label{fig:jplot}
\end{figure}

The 15 CMEs listed in Table~\ref{tab:events} are independently tracked by three scientists (S1, S2, S3), each with extensive experience in analysing HI data. All use the same data reduction pipeline described in Section~\ref{sec:data}. Therefore, any differences between the resulting time-elongation tracks arise solely from individual judgment in identifying the CME leading edge.
The tracking tool initially displays only a subset of the available HI data. The tool allows the user to adjust the contrast of the Jmap to their convenience. After completing five manual measurements of the CME front, the user advances the sequence by clicking a button that reveals the next four hours of data. The scientist then reperforms another five measurements of the same CME front in this extended time range to obtain a new set of five tracks. This process is repeated until the maximum track length is reached, that is, until the CME can no longer be followed. As an example, the upper panel of Figure~\ref{fig:jplot} shows such a Jmap for CME15 for the minimum amount of data for this event and the lower panel illustrates the maximum track length for CME15, along with the corresponding mean tracks from each scientist.

\subsection{Graduated Cylindrical Shell Fitting}
\label{sec:GCS}

\begin{figure}[h!]
\centering
\includegraphics[width=0.8\textwidth]{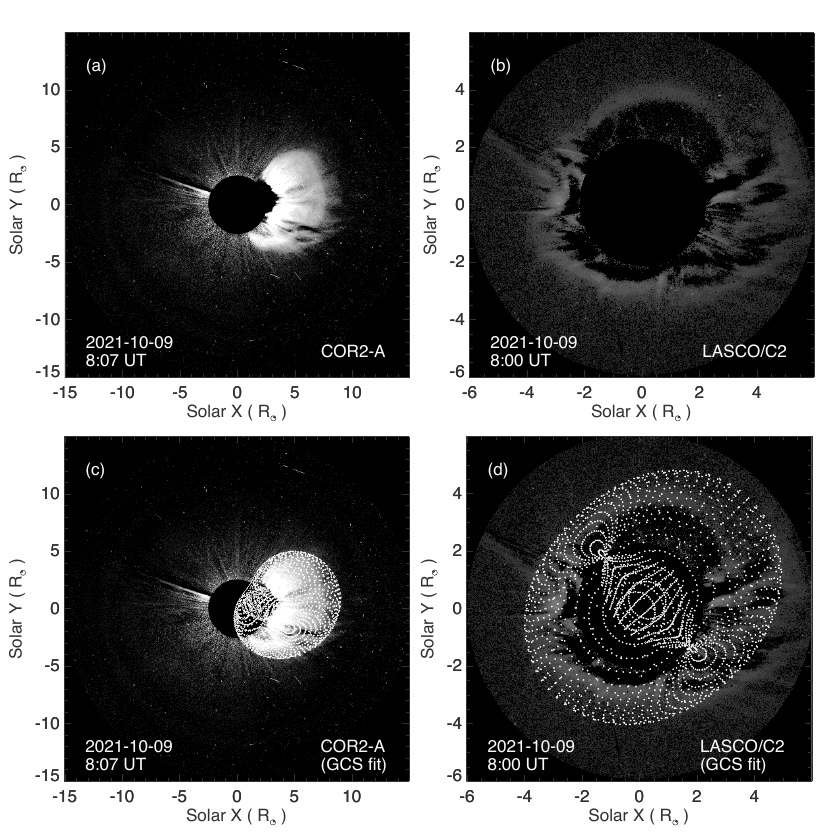}
\caption{Panels a) and b) show snapshots of the CME that occurred on 9 October 2021 in (a) STEREO-A/COR2 and (b) SOHO/LASCO-C2 FOV view. Panels c) and d) show the CME fitted with the GCS model, with the white wireframe structure generated from the fit overlaid on the respective coronagraph images.}
\label{fig:gcs}
\end{figure}

To reconstruct the 3D structure of CMEs, Level 0.5 data of EUVI, COR1 and COR2 were processed to Level~1 using \texttt{secchi\_prep.pro}, and Level 0.5 data of SoHO/LASCO-C2/C3 was processed to Level~1 using \texttt{reduce\_level\_1.pro} in IDL. Next, base difference images were created by subtracting a pre-CME image from successive images. Panels a) and b) in Figure~\ref{fig:gcs} show snapshots of CME15 that occurred on October 9, 2021 in the STEREO-A/COR2 and LASCO-C2 FOV. Then, the Graduated Cylindrical Shell model \cite<GCS;>{the09} was applied to observations from the SOHO/LASCO-C2/C3 coronagraphs and the COR2 coronagraphs on-board either of the STEREO-A (for CMEs occurring from 2020 onwards) or STEREO-B (for the two CMEs occurring in 2010, see Table~\ref{tab:events}) spacecraft. This is a realistic set-up for CMEs that could be observed by Vigil together with a coronagraph near Earth. Panels c) and d) in Figure~\ref{fig:gcs} present an example of the GCS hollow croissant fitted to the same CME . The fitting procedure that was followed is outlined in \citeA{the09} and \citeA{maj20} and was repeated for the different timestamps, for which the CMEs were observed in the combined FOV of COR2 and LASCO-C2/C3 to capture their 3D evolution.

The GCS model has six free parameters for fitting, but using only two vantage point observations can often result in degeneracies in some of the geometrical parameters like tilt-angle, half-angle, longitude, latitude etc. \cite<see>{the09, maj22}. In such cases, it is often advisable to constrain some of the model parameters. Thus, in this work the source regions of the CMEs were identified using JHelioviewer \cite{mul17} by backtracking the CMEs onto the solar disk, where the source regions associated with the CMEs were identified using the disk imaging observations from the various passbands of SDO/AIA and STEREO/EUVI. For details on source region identification using JHelioviewer, please refer to \citeA{maj23}. The locations of the source regions were then used to better constrain the latitude and longitude of the GCS shape. However, for the cases when a CME has the same appearance from an axial or lateral perspective from both vantage points \cite{cre20} a third vantage point plays a crucial role in estimating the tilt-angle and the half-angle parameters \cite{maj22}. Since in this work, the third vantage point is not used, the angular width calculated from the GCS model might still be an underestimate of the actual angular width of the CME.

In the following analysis, for the case where we use GCS input to ELEvoHI, we use the GCS longitude as input for the CME apex propagation direction and the GCS face-on width as input for the angular width in the ecliptic plane (see Section~\ref{sec:elevohi}). The GCS face-on width, $f_{\mathrm{w}}$, is calculated using the following equation:

\begin{equation}
    f_{\mathrm{w}} = 2(\alpha + \sin^{-1}k),
\end{equation}

where $\alpha$ is the GCS half-angle and $k$ is the aspect ratio \cite{maj22}.

\subsection{Fixed-Phi Fitting}

Fixed-Phi Fitting \cite<FPF;>{rou08} is one of the simplest methods using HI data for modelling and predicting CME arrival times. The method fits a function to the HI time-elongation track from a single spacecraft by assuming a point-like CME shape and a constant propagation speed. Both assumptions are considerable oversimplifications as a CME is a huge structure that interacts with the ambient solar wind on its way through the heliosphere. However, a constant speed within the HI FOV can occur for events that have approximately reached the same speed as the surrounding medium. This is the case for slow CMEs that are accelerated by the drag-force exerted by the solar wind and reach their final propagation speed in the first half of the HI1 FOV or for fast CMEs undergoing a strong deceleration in their early evolution \cite{gop00}. In this context, previous studies have found that the heights at which drag forces take over during a CME's journey lie predominantly in the middle corona \cite{sac17}. Thus, it seems a reasonable assumption to consider uniform CME speeds at heliocentric distances imaged by heliospheric imagers.

FPF was among the first methods developed to utilise HI data in the beginning of the STEREO era. Other HI time-elongation fitting methods have different assumptions regarding CME shape and width but also presume constant propagation speed. For information on a method assuming a constant CME width of 180$^\circ$ (Harmonic Mean Fitting, HMF) see \citeA{lug09a}, and for a method permitting selection of any width from 0 to 180$^\circ$ (Self-Similar Expansion Fitting, SSEF) see \citeA{dav12}. 

Besides generating a prediction of the arrival time at 1 au, FPF is a fast and easy tool to get an estimate of the approximate CME speed and the direction of motion. In the following analysis, for the case where we use FPF input to ELEvoHI, we use the FPF-derived longitudinal direction and the angular half width is assumed to be the same for all events, namely $45 \pm 10^\circ$.

\subsection{ELEvoHI in brief}
\label{sec:elevohi}

The ELEvoHI modelling framework for CME arrival predictions constitutes a significant advance on the single-spacecraft geometrical fitting of HI time-elongation profiles discussed above as it also accounts for the interaction of the CME with the ambient solar wind. This is done through the inclusion of an elliptical CME front to better capture the expected CME distortion, and the decelerating/accelerating effect of the solar wind through a drag-based equation of motion. A detailed description of the multi-stage ELEvoHI procedure is provided by \citeA{rol16} (single-run mode) and \citeA{ame18} (ensemble mode). \citeA{ame21} provides an informative introduction to the ELEvoHI methodology, including a schematic of the different elements that make up the forecasting scheme. Further refinements, but not used in this study, such as incorporating a deformable front can be found in \citeA{hin21a} and \citeA{hin21b}.

Like for the other HI fitting methods, the primary input to ELEvoHI is the time-elongation profile of the CME front taken from a single HI vantage point. An additional commonality between ELEvoHI and these methods is the assumption of self-similar expansion of the CME, predicating an invariant cross-sectional angular half width of the CME's assumed elliptical form, along a fixed apex direction.

Once a HI time-elongation profile is acquired, the first step in ELEvoHI is to convert that profile, via the ELlipse Conversion module \cite<ELCon;>{rol16}, into an equivalent radial distance profile. It is assumed that the line-of-sight from the observer --- corresponding to an observed elongation --- forms the tangent to the leading edge of the CME. The conversion proceeds by means of inputting an apex propagation direction, half width and inverse aspect ratio of the ellipse half-axes. As discussed in the aforementioned publications, ELEvoHI generally performs this conversion by using the apex direction obtained from one of two different methodologies (GCS and FPF, see above) as input.

\begin{figure}
\includegraphics[width=\textwidth]{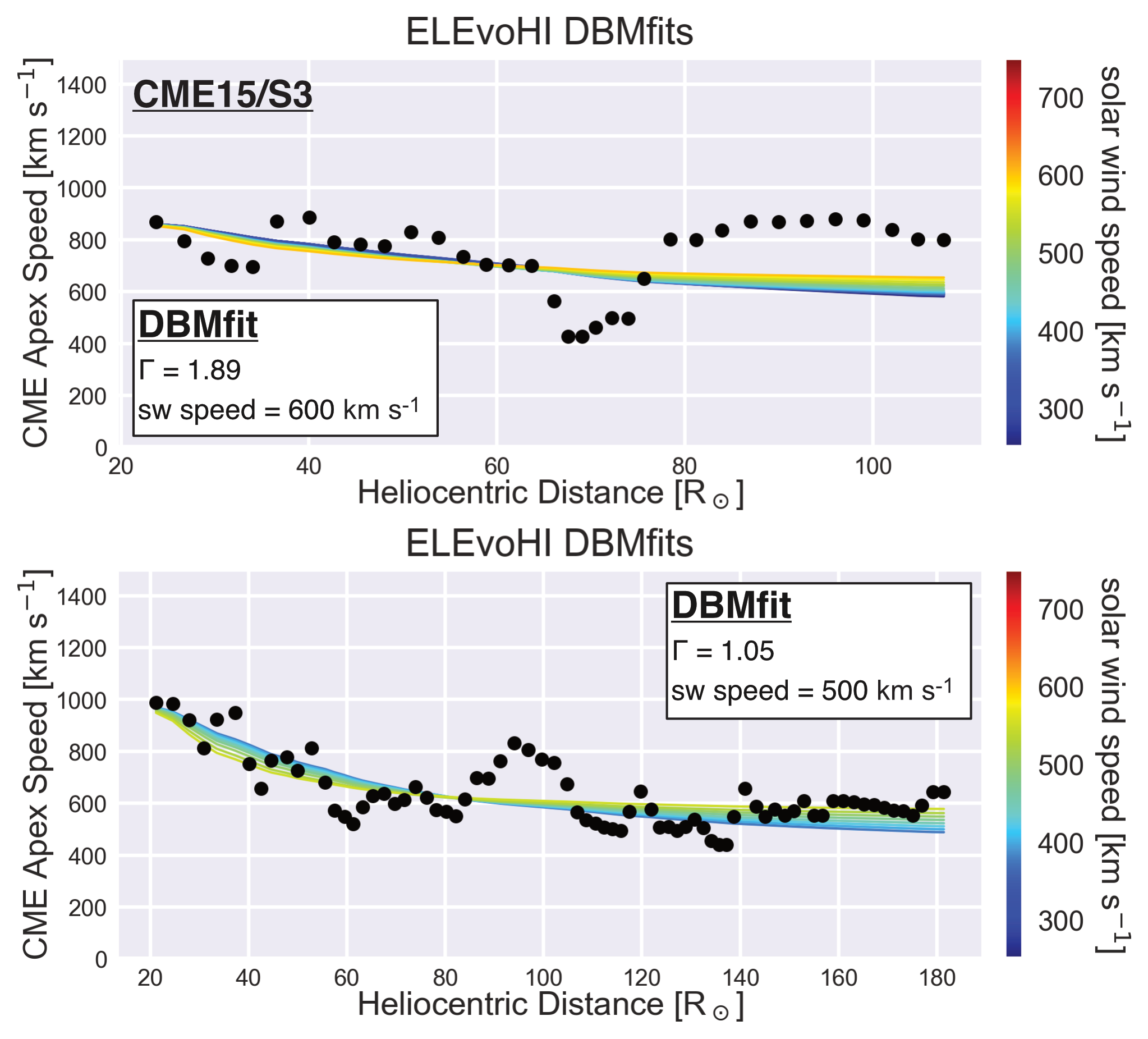}
\caption{\small DBMfitting applied to two different track lengths of the same CME. The black dots represent the CME kinematics derived using ELCon based on the direction calculated using FPF. The coloured lines are the different DBMfits corresponding to several possible combinations of solar wind speed (colour-coded) and drag parameter ($\Gamma$). Although both tracks are gained from the same event and the direction output from FPF ($\phi$) differs only by $5^\circ$, the resulting CME kinematics diverge, leading to a different ambient solar wind speed derived by DBMfitting and subsequently to a different ELEvoHI arrival prediction.} 
\label{fig:DBMfit}
\end{figure}

The radial distance profile and radial speed profile derived by differentiation thereof, that are output by ELCon, are required input for the subsequent component in the ELEvoHI CME forecasting framework, Drag-Based Model fitting \cite<DBMfitting;>{rol16}. Here, the radial distance profile from ELCon is fitted using a drag-based equation of motion \cite{car04} utilised in the drag-based model \cite<DBM;>{vrs13}. DBM considers the influence of the drag force acting on solar wind transients during their propagation through interplanetary space, and is based on the assumption that, beyond a distance of about 15 R$_\odot$ from the Sun, the driving Lorentz force can be neglected and the drag force can be considered as the predominant force affecting the propagation of a CME. The drag imposed on a CME not only depends on the instantaneous CME speed, but also on the corresponding ambient solar wind speed and a so-called drag parameter. In the DBMfitting procedure, the best fit (i.e.\ smallest mean absolute residual) of the drag equation to the radial distance profile from ELCon provides, as output, estimates of the mean solar wind speed and drag parameter, both of which are assumed to be constant over the extent of the CME track. Limitations are set on the possible range for each of these parameters within DBMfitting, such that they remain physically realistic. For the solar wind speed, we consider values between 250 and 775~km s$^{-1}$, and we limit the drag parameter to a maximum of $2 \times 10^{-7}$~km$^{-1}$.

For the final component of ELEvoHI, the ELlipse Evolution model \cite<ELEvo;>{moe15}, uses the best-fit output of the DBMfitting procedure and generates a predicted CME arrival time and speed at any specified target in the heliosphere --- in this study Earth --- taking into account the curvature of the CME front modelled with an ellipse shape. Via ELEvo, ELEvoHI can generate arrival time predictions beyond the extent of the input HI time-elongation profile in 2D along the modelled CME front.  

In ensemble mode, ELEvoHI is run by specifying a number of combinations (generally several hundred) of the inverse ellipse aspect ratio, ellipse angular half width (both within the ecliptic plane), and apex propagation direction. 
The range of each parameter is defined relative to the input value that is either obtained from FPF or GCS, or assumed canonically. From those ensemble members that are predicted to impact the target, the median arrival time and speed are derived, which can be compared with in situ values to assess the accuracy of the prediction. Additionally, we examine the prediction results based on the exact input parameters (usually the middle member within the ensemble) and name this prediction deterministic run.

Although previous studies \cite{ame18,ame21,hin21a,hin21b} used the angular width derived from the intersection of the GCS croissant with the ecliptic plane as input to ELEvoHI, this study simply uses the GCS face-on width as input for the width within the ecliptic plane.
\citeA{palkay24} found a standard deviation of 8$^\circ$ for the longitude and $9.3^\circ$ for the angular width. Therefore, we build the ensemble within ELEvoHI by varying the propagation direction and the angular half width by $\pm 10^\circ$, with a step size of 2 and $5^\circ$ respectively, and the inverse aspect ratio between 0.7 (flatter front) and 1 (circular), with a step size of $0.1$, leading to an ensemble size of 220 ($11 \times 5 \times 4$) members. 

\section{Data assimilation approach}

To simulate a real-time environment for the future Vigil mission that is going to be located around the L5 point and will observe Earth-directed CMEs from a side view, it is assumed that in an operational setting more and more data will be received as time progresses and can then be included in the prediction. One might presume that the Earth arrival prediction would improve with the utilisation of more observational data from the CME evolution. In order to test this hypothesis, we apply ELEvoHI repeatedly to progressively longer extents of the time-elongation track of each of the 15 selected events. 

For running ELEvoHI, we at least need a propagation direction and the time-elongation track itself as input. More information about the shape of the CME, e.g.\ the half width of the intersection with the ecliptic plane or the curvature of the front, can be included too as shown in \citeA{ame21}. As described above, for deriving the direction we use two different sources, the FPF method and GCS fitting, the latter applied to coronagraph data. When FPF is used, we assume the angular half width to be the same for each event, namely $45 \pm 10^\circ$. When GCS is used, we assume the angular width to be the same as the GCS face-on width with an uncertainty range of $\pm 10^\circ$. Since GCS is not using the HI time-elongation track, the GCS-direction does not depend on the length of the track and stays the same for each prediction of an event. Contrary to that, FPF is applied to the same track length as used subsequently by ELEvoHI, hence the FPF-direction is a function of track length (and tracking scientist) and additionally influences the accuracy of the prediction. However, GCS fitting was only possible for 9 out of the 15 events. The reasons for this are either data gaps or a too faint structure to be fitted in coronagraph data.

Figure \ref{fig:DBMfit} shows how the data assimilation is implemented within ELEvoHI for one example event. The upper panel shows the CME kinematics resulting from the ELCon elongation-to-distance conversion method (black dots) based on the FPF-direction and applied to one of the shorter time-elongation track available for this event (max. elongation is $31.3^\circ$).
The colour-coded lines are the DBMfits performed for different ambient solar wind speeds.
The lower panel is an example for the maximum track length available for the same event (max. elongation is $56.5^\circ$). Applying the FPF method to tracks with a different length yields different propagation directions, leading to distinct CME kinematics (including a different initial speed) compared to the upper panel. Consequently, the solar wind speeds derived from DBMfitting also differ. Together, these variations produce a different prediction for the CME's arrival time and speed at Earth.

In order to investigate the influence of the person manually extracting the time-elongation measurement points from the time-elongation map, we conduct the approach described above for each of the three scientists, S1, S2, and S3, separately.

In the following section we explore the findings of this strategy and try to find a best practice set-up to potentially use future Vigil data in a real-time prediction setting.

\section{Results}

\begin{figure}
\includegraphics[width=\textwidth]{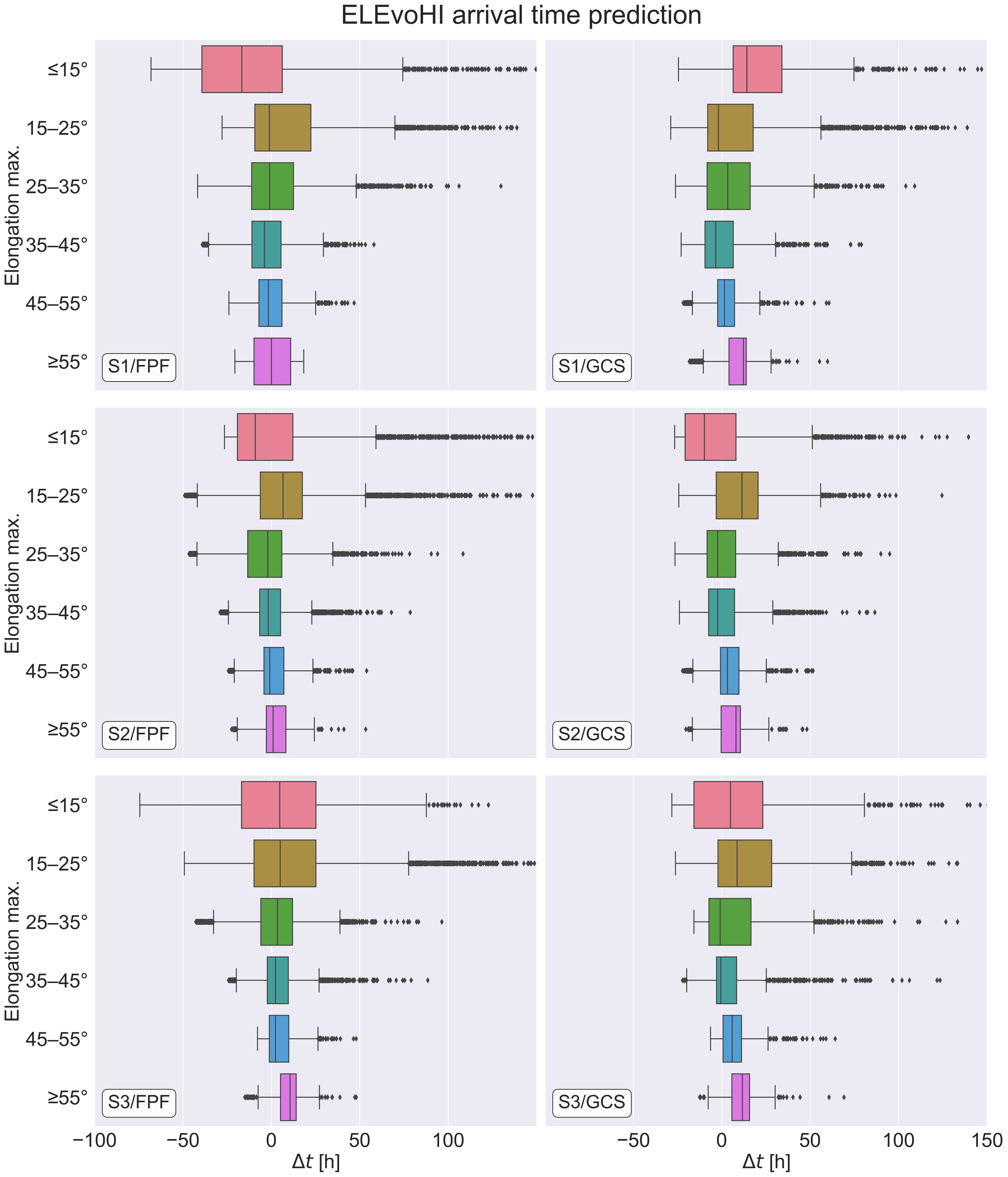}
\caption{\small Difference of modelled and in situ detected arrival times, $\Delta t$, based on scientist tracking (S1, S2, S3), maximum elongation (track length) used and method to derive the direction of motion (FPF and GCS). Positive (negative) $\Delta t$ means the arrival was predicted too late (too early).} 
\label{fig:elevohi_all}
\end{figure}

Figure \ref{fig:elevohi_all} shows the ELEvoHI ensemble results related to FPF (left column) and GCS (right column), S1 (upper row), S2 (middle row) and S3 (lower row). Plotted are the differences between modelled arrival time and in situ detected arrival time ($\Delta t$ in hours) for each CME for a binned maximum elongation range. The boxes encompass 50$\%$ of the data, the vertical solid line within the boxes is the median value and the horizontal whiskers extend from the boxes to the smallest and largest values within 1.5 times the interquartile range, the diamonds are outliers. The number of modelled CMEs is significantly lower when GCS is used. The reason for this is that we choose the set of events based on good quality HI data and in situ observations rather than based on a good visibility in coronagraph data. This leads to only 9 out of the 15 events that can be fitted using GCS (marked with an asterisk in Table~\ref{tab:events}). 

Overall, the median $\Delta t$ is closer to $\Delta t = 0$, i.e.\ the exact match of predicted and observed arrival time, when more HI data (longer time-elongation tracks) is used. This tendency is present for both methods (FPF or GCS) used to derive the CME direction. 
Table~\ref{tab:metrics} lists the metrics (averaged over the results of S1, S2, and S3) for ELEvoHI/FPF and ELEvoHI/GCS and the different maximum elongations used. As a benchmark, we state the metrics for the available CME Scoreboard events (compare Table \ref{tab:events}) in the last row. Based on that, our results show that including HI data at least out to $35^{\circ}$ elongation improves the mean absolute error of the arrival time predictions by about 3 hours compared to forecasts not based on HI data (as on CME Scoreboard). The numbers in brackets represent the metrics for only those events that were modelled by ELEvoHI/FPF and ELEvoHI/GCS as well as included in the CME Scoreboard. The events for which all three data sources provide results are marked with an asterisk and dagger in Table~\ref{tab:events}. This approach leaves only six events, but reveals that the differences between ELEvoHI/FPF and ELEvoHI/GCS become minimal under consistent event coverage. The comparison with the benchmark remains unchanged, indicating that predictions incorporating HI data out to at least $35^\circ$ continue to outperform those that do not include HI data at all.

Differences in the modelling results based on the several approaches are discussed in detail below.

\begin{table}[htbp]
\centering
\caption{Arrival time prediction metrics (MAE: mean absolute error, ME: mean error, RMSE: root mean square error, STD: standard deviation) of ELEvoHI/FPF and ELEvoHI/GCS based on the time-elongation tracks of S1, S2, and S3. As a benchmark, metrics from CME Scoreboard derived from the same events as available (marked with a dagger in Table~\ref{tab:events}) are given. The numbers in brackets are calculated from the common events (marked with asterisk and dagger in Table~\ref{tab:events}). Values are given in hours.}
\label{tab:metrics}
\begin{tabular}{lcccc}
\hline
ELEvoHI/FPF &  &  &  &  \\
\hline
Max. &  &  &  &  \\
Elongation & MAE & ME & RMSE & STD \\
\hline
$\leq15^\circ$ & 27.8 (27.6) & -2.0 (19.4) & 36.8 (38.8) & 35.9 (32.8) \\
$15-25^\circ$ & 19.6 (18.7) & 7.9 (4.9) & 27.3 (24.6) & 26.1 (23.3) \\
$25-35^\circ$ & 13.3 (12.6) & -0.2 (4.3) & 17.3 (15.9) & 17.1 (15.3) \\
$35-45^\circ$ & 9.2 (8.5) & -0.2 (3.6) & 12.4 (11.6) & 12.0 (10.7) \\
$45-55^\circ$ & 7.6 (7.1) & 0.7 (6.1) & 10.2 (9.6) & 9.7 (7.3) \\
$\geq55^\circ$ & 9.2 (9.6) & 3.6 (9.3) & 11.0 (10.7) & 9.3 (4.7) \\
\hline
ELEvoHI/GCS &  &  &  &  \\
\hline
Max. &  &  &  &  \\
Elongation & MAE & ME & RMSE & STD \\
\hline
$\leq15^\circ$ & 22.9 (27.5) & 9.4 (21.2) & 30.4 (36.2) & 27.8 (27.6) \\
$15-25^\circ$ & 17.0 (17.2) & 10.3 (13.9) & 23.8 (23.4) & 21.2 (18.2) \\
$25-35^\circ$ & 12.9 (14.1) & 4.1 (11.0) & 17.3 (19.6) & 16.7 (16.1) \\
$35-45^\circ$ & 8.5 (7.6) & 1.5 (4.9) & 12.3 (12.5) & 12.2 (11.5) \\
$45-55^\circ$ & 7.6 (6.8) & 4.0 (6.3) & 10.4 (9.5) & 9.3 (7.2) \\
$\geq55^\circ$ & 11.1 (11.4) & 8.5 (11.2) & 12.9 (12.8) & 9.4 (5.8) \\
\hline
Benchmark &  &  &  & \\
CME Scoreboard & 12.3 (13.7) & -5 (-8.4) & 16.6 (18.1) & 15.9 (16.1) \\
\hline
\end{tabular}
\end{table}

\subsection{Influence of the person tracking}

To compare the tracks from S1, S2, and S3 we plotted them on each of the Jmaps to visualise the differences. Figure \ref{fig:jplot} is a representative example of what this comparison looks like. For all events, the tracks are well matched, but sometimes with differences in the achieved track length. The average track length for S1 (S2, S3) is $47.2^\circ$ ($47.6^\circ$, $46.3^\circ$) with a standard deviation of $12.1^\circ$ ($9.5^\circ$, $12.3^\circ$).
A link to a comparison of all tracks for the whole event list can be found in Section~\ref{sec:data_access}.
However, despite the overall agreement of the tracked feature and the overlapping tracks, the results based on one scientist or the other differ for some events by quite a large amount. This leads to large differences of the metrics related to ELEvoHI predictions using the shortest time-elongation tracks. Tracks with a maximum elongation up to $15^\circ$ from S1 (S2, S3) result in a mean absolute error, MAE($\Delta t$), of $32.2$ ($23.7$, $27.5$) hours (ELEvoHI/FPF) and $25.3$ ($20.4$, $23.0$) hours (ELEvoHI/GCS) and a bias-related mean error, ME($\Delta t$) of $-13.6$ ($4.3$, $3.5$) hours (ELEvoHI/FPF) and $20.2$ ($-0.4$, $8.4$) hours (ELEvoHI/GCS).
All these metrics improve with growing track length (mean over all scientists is listed in Table \ref{tab:metrics}).
The best accuracy is reached using a track length between 45 and $55^\circ$. S1 (S2, S3) achieves an MAE($\Delta t$) of $7.9$ ($8.5$, $6.5$) hours (ELEvoHI/FPF) and $7.2$ ($8.4$, $7.3$) hours (ELEvoHI/GCS) with an ME($\Delta t$) of $-1.7$ ($-1.0$, $4.9$) hours (ELEvoHI/FPF) and $1.4$ ($3.7$, $6.9$) hours (ELEvoHI/GCS).
When more HI data, i.e.\ beyond $55^\circ$ elongation, is included, the prediction accuracy decreases again.

\begin{figure}[h!]
\includegraphics[width=\textwidth]{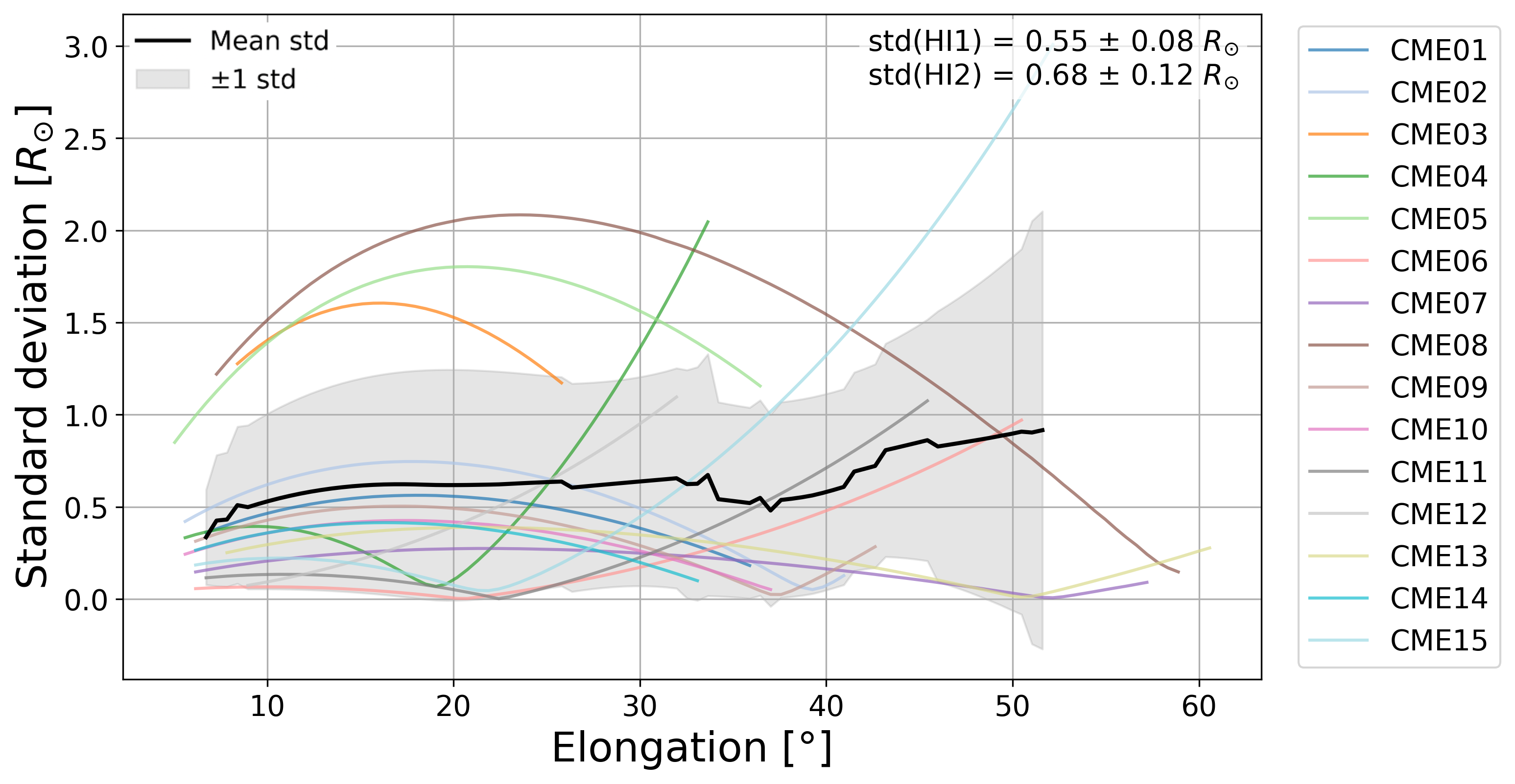}
\caption{\small Standard deviation of the reconstructed CME apex in units of solar radii, shown as a function of elongation for all events and averaged over the tracks produced independently by the three scientists. Each coloured curve represents one CME, while the black line shows the mean standard deviation across all events. The grey shaded area denotes $\pm 1$ standard deviation. For the conversion from elongation to heliocentric distance ELCon based on the FPF-direction was used.}
\label{fig:std_tracking}
\end{figure}

Usually, there are several possibilities for large differences of the results from different persons tracking. First, the Jmap that is used could be different. Typically, scientists use their own procedures to gather the measurements of the CME fronts. Although in this study all scientists use the same tracking tool, this possibility cannot be completely ruled out, as they are offered the option to adjust the contrast in the Jmaps to their own preference. Second, the understanding of where to put the measurement points is different, which seems to be the main reason for the differences here. Figure \ref{fig:std_tracking} shows the standard deviation of the ELCon CME kinematics in solar radii (R$_\odot$) based on the longest tracks of each scientist for the whole set of CMEs and the corresponding deterministic run based on ELEvoHI/FPF. Within the HI1 FOV, i.e.\ below 18$^\circ$ in our case because in the overlap region we used HI2 data (compare Figure~\ref{fig:jplot}), the tracking error is $0.55 \pm 0.08$~R$_\odot$ and within the HI2 FOV the tracking error is $0.68 \pm 0.12$~R$_\odot$.

Still, although the tracks do generally agree with each other, the resulting predictions show a relatively wide range. As is obvious from this study, even small differences in the tracks can lead to large differences in the prediction result. But there is a lot of potential for unifying our tracking approaches leading to a more stable and hopefully more accurate prediction.
The Solar Transient Recognition Using Deep Learning \cite<STRUDL;>{bau25} for HI data is an ideal possibility to minimise the influence of manually tracking a CME front. STRUDL automatically detects and outlines CMEs in HI images. It can also follow the CME fronts from one image to the next, creating time-elongation profiles that can further be used by ELEvoHI to perform the arrival forecast. 
Another possibility is a tracking tool that allows to view a CME not only in the time-elongation map (compare Figure~\ref{fig:jplot}) but simultaneously in the according HI images. Such a tool is already available and was used and presented in \citeA{lel25}.

\subsection{Influence of direction method}

The width and longitude (and in 3D also the latitude) of a CME is the only set of parameters that determines if an event is going to hit the target or miss it. Currently, all operational tools that model the CME evolution assume these attributes to stay constant from the coronagraph FOV onwards, i.e.\ they assume self-similar expansion.
This roughly seems to be a fair conjecture as changing the direction of such an extended feature like a CME might be only possible close to the Sun due to rotation of the whole structure or due to deflection by e.g.\ a coronal hole \cite{moe15,kay17,sah20,wan20}. However, the high rate of false alarms (predicted to hit but in fact missed), namely around $40\%$ \cite{sco18}, results either from erroneous parameters derived from coronagraph observations or from a non self-similar evolution. Due to a lack of suitable imaging observations from various vantage points to pin down the CME 3D shape we cannot prove the latter assumption. The PUNCH mission may be able to help us refine our understanding of CME expansion during propagation as it provides an additional HI vantage point.

The discussion above applies not only to GCS that is applied to coronagraph observations (and similar tools as the SWPC-CAT or STEREO-CAT), but also to HI-based tools to derive the direction of the CME, such as FPF or SSEF. In addition to the self-similar evolution, these methods assume a constant propagation speed to estimate the direction. However, they utilise imaging data from larger heliospheric distances and might be capable of taking into account influences the CME undergoes during its evolution to a certain degree. A comparison of GCS and FPF as a source of propagation direction (and in the case of GCS also of the width) as input for ELEvoHI was already performed in \citeA{ame21} and showed a slight improvement of the prediction accuracy when using GCS.

 As already explained above, we compiled the list of CMEs based on good HI and in situ observations and tried to find CME-ICME pairs that could be associated close to certainty. Due to this approach, coronagraph data was not usable for all events under study because the front was either too faint or not visible at all. For the 9 events left from our list, we see that if coronagraph data can be used, including them could lead to an improvement for certain events if the availability of HI data is limited to below $25^\circ$. If we only compare the results for events that are modelled by both ELEvoHI/FPF and ELEvoHI/GCS (events marked with an asterisk in Table~\ref{tab:events}) the differences in the metrics become negligible and with an MAE($\Delta t$) of $22.3$ hours for elongations below $15^\circ$ even smaller for ELEvoHI/FPF than for ELEvoHI/GCS. Moreover, using HI observations as the basis for real-time predictions may help identify more Earth-directed events that could otherwise be missed when relying solely on coronagraph data \cite{har23}.

\subsection{Influence of the maximum elongation}

\begin{figure}[h!]
\includegraphics[width=\textwidth]{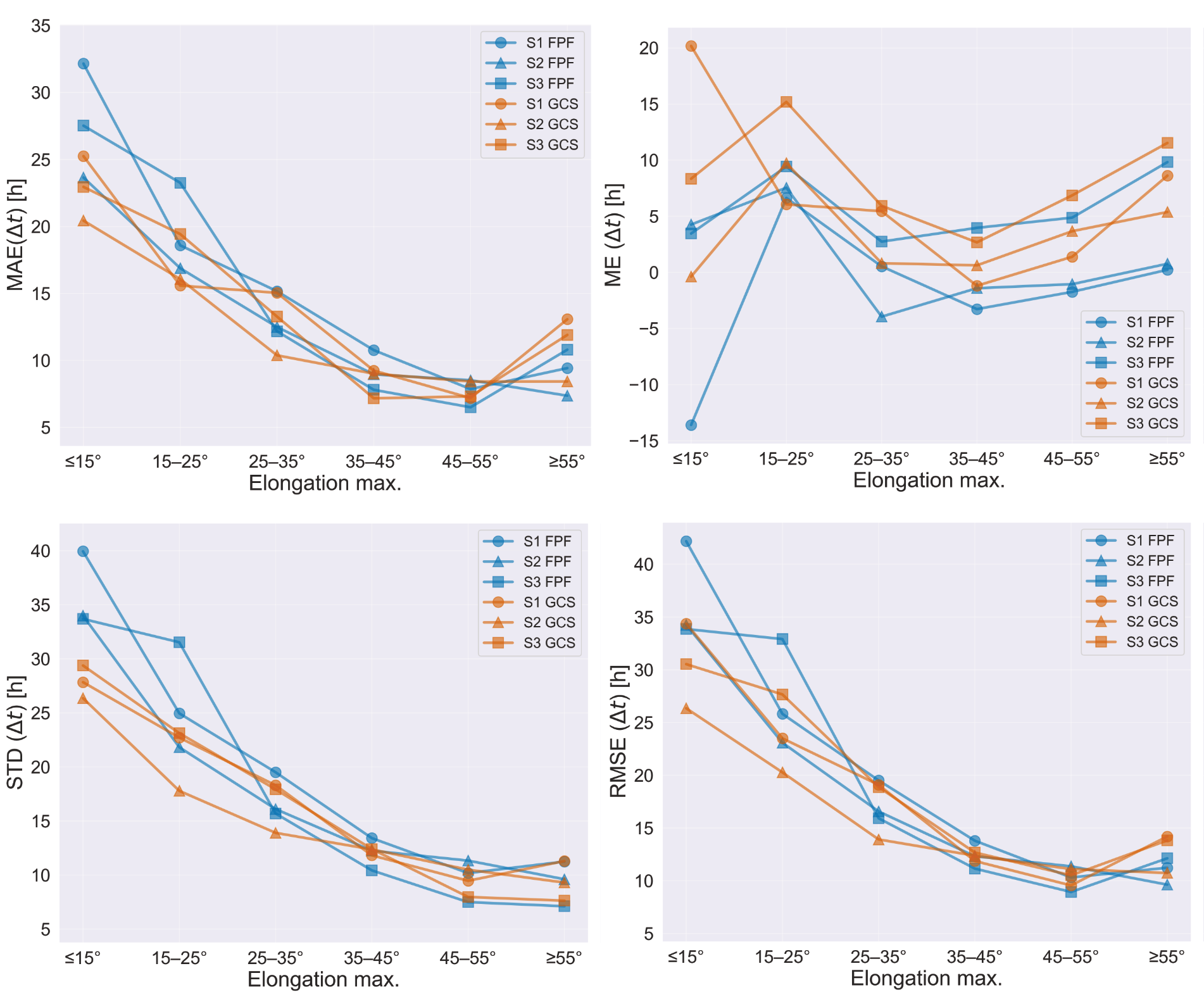}
\caption{\small MAE($\Delta t$) (upper left), ME($\Delta t$) (upper right), STD($\Delta t$) (lower left) and RMSE($\Delta t$) (lower right) based on ELEvoHI/FPF (blue) and ELEvoHI/GCS (orange) and the three sets of tracks from different scientist (S1: circle, S2: triangle, S3: square) as a function of maximum elongation used.}
\label{fig:metrics_all}
\end{figure}

Figure \ref{fig:metrics_all} shows the ELEvoHI model metrics based on the different tracks and the CME direction from FPF (blue) and direction/width from GCS (orange). The effect of the track length is similar across the three scientists. For all scientists and both methods, the prediction accuracy improves as the track length increases. Notably, the MAE($\Delta t$) exhibits a relatively large spread for short tracks, but this spread decreases as the tracks grow longer, nearly overlapping for track lengths between $35^\circ$ and $55^\circ$ elongation.

For tracks longer than approximately $35^\circ$ the MAE($\Delta t$) improves over the CME Scoreboard. However, at a separation of around $60^\circ$ between the remote observer and the in situ detection (as it is planned for Vigil), an elongation value of $~55^\circ$ is already close to impact, i.e.\ we need to take into account that such observations cannot be used for real-time prediction with a sufficient prediction lead time.

For some events it is possible to improve the prediction when the first measurement points are not considered for modelling. Cutting off the earliest points of the time-elongation track can even change a result from a non-converging DBMfit, i.e.\ no prediction possible, to a good prediction result. The reason for this is likely an unusual behaviour in the early evolution phase, such as ongoing acceleration. This could either be a true behaviour of the CME or an artefact from the overexposure in the early HI1 FOV. The same can happen in the beginning of the HI2 FOV and can even be enhanced due to the underexposure in the late HI1 FOV. When doing case studies it is easier to examine the CME kinematics in detail and allow only those parts to contribute to the time-elongation profile that are doubtlessly correct. However, in this study, we treat every CME in the same way and include the beginning of the tracks of each event. In the future it might be a feasible approach to vary the beginning of the track as an additional ensemble parameter, therefore avoiding situations where no DBMfit is possible due to late acceleration of the CME in the early HI1 FOV.

\begin{figure}[h!]
\includegraphics[width=\textwidth]{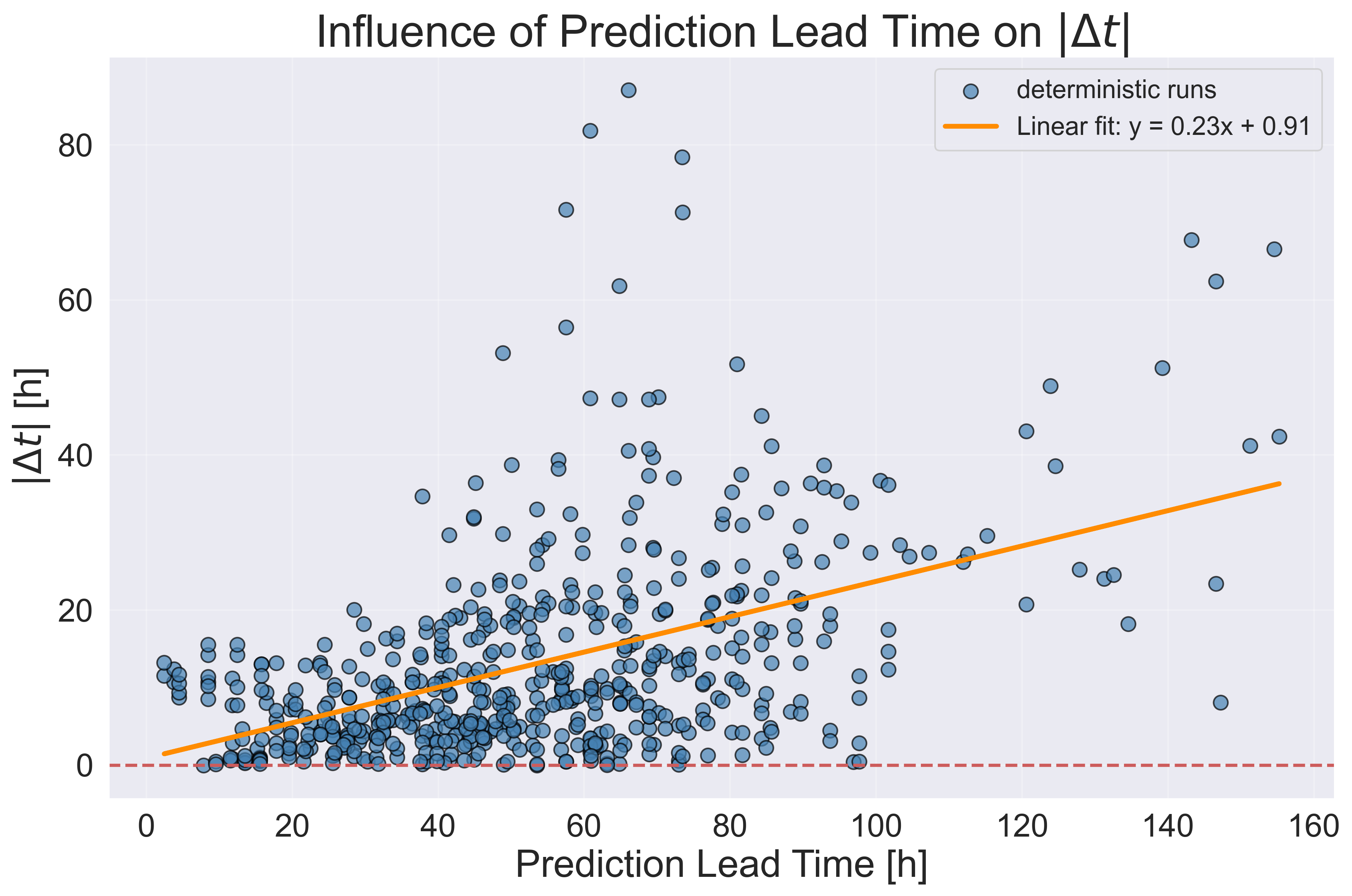}
\caption{\small Absolute time difference between predicted and observed arrival time as a function of prediction lead time based on the deterministic runs from all available tracks.} 
\label{fig:PLT}
\end{figure}

Although using additional HI data improves the prediction accuracy over time, it also reduces the forecast lead time, leaving less time to prepare for a potentially geoeffective event. In general, the prediction lead time is defined as the interval between when a forecast is issued and when the CME reaches the target. In this study, we consider the time between the last HI data point used and the CME's in situ detection. Figure \ref{fig:PLT} shows the prediction lead time for all deterministic runs (586 that are predicted to hit Earth) in our sample ($127361$ single runs in total), along with the corresponding absolute $\Delta t$. The fitted relationship suggests that the absolute arrival time error decreases by about 0.23 h ($\sim$14 min) for every hour by which the prediction lead time is shortened. Expressed in relative terms, reducing the lead time from 100 h to 50 h corresponds to an improvement in forecast accuracy of roughly 45 \%.

\subsection{Influence of ensemble building}

We build our ensemble in ELEvoHI simply by varying the shape-related input parameters according to the commonly known accuracy of GCS reconstruction. Since we do not weight the ensemble, every ensemble member adds to the result with the same likelihood. However, it is safe to assume that the actual derived GCS parameters (or FPF direction) is more likely than those at the borders of the parameter space. Usually, the median of the ensemble is used as arrival time (and arrival speed) when stating the prediction, while the error is given by the standard deviation within the ensemble.

\begin{figure}[h!]
\includegraphics[width=\textwidth]{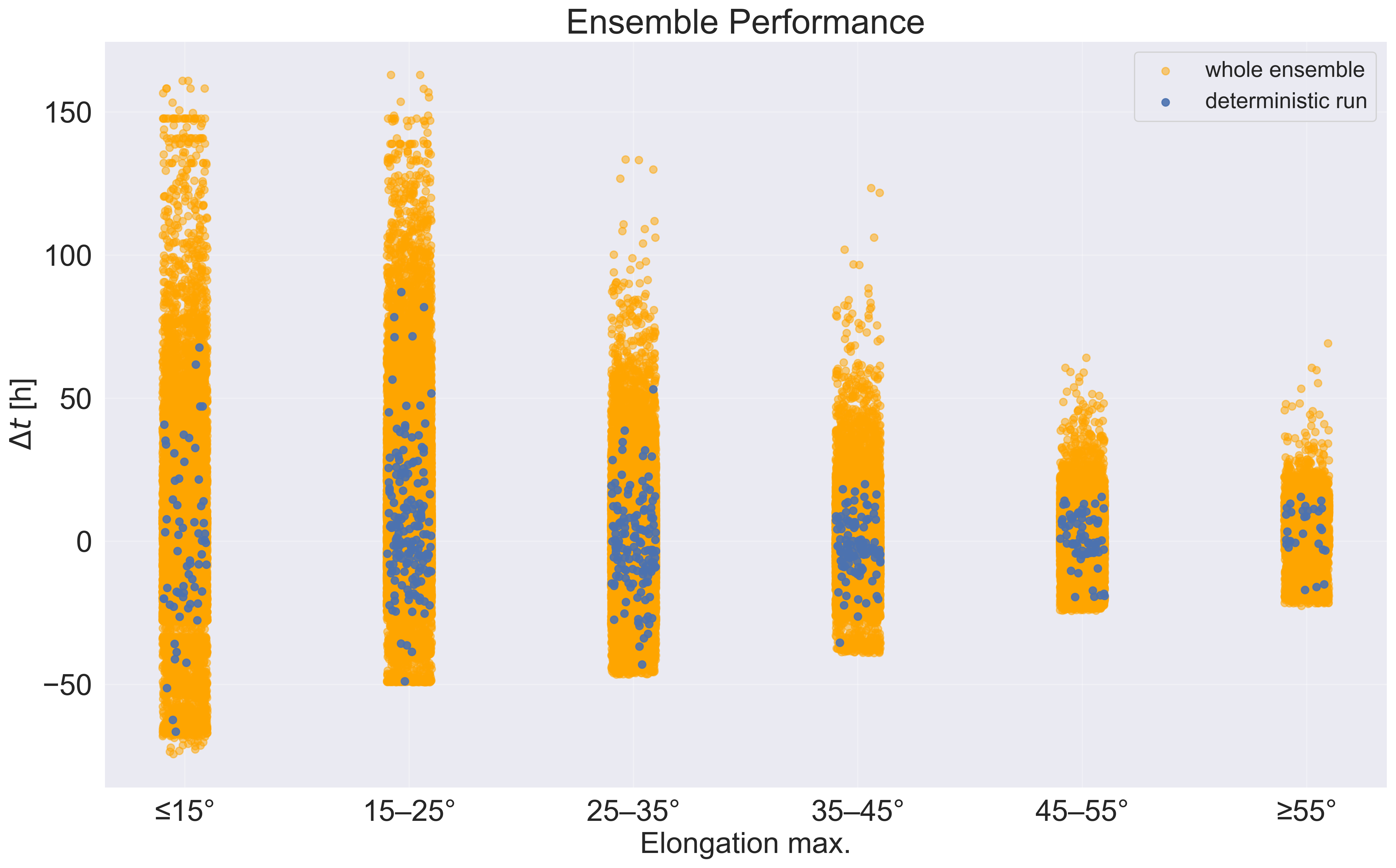}
\caption{\small Difference between predicted and observed arrival time ($\Delta t$) of all 15 events of each ensemble member (orange dots) and each deterministic run (blue dots) based on the maximum elongation values used.} 
\label{fig:deterministic}
\end{figure}

To test if weighting the ensemble improves the forecast we look at the result of the deterministic run, i.e.\ the run based on the actual results of GCS (or FPF) and the middle member of the ensemble. Figure \ref{fig:deterministic} represents $\Delta t$ of the whole ensemble (orange dots) and of the deterministic runs (blue dots) for the whole set of 15 events. For our event sample, deterministic runs bring better results than the whole ensemble. Therefore, building asymmetric errors around the deterministic run rather than using the ensemble median or mean with symmetric error ranges might improve the prediction accuracy and offer a more realistic representation of possible outcomes. 

It is also worth noting that some directions lead to highly unlikely CME kinematics, which are currently nonetheless used and included in our ensemble. In the future, we will rethink our method to build the ensemble and hope that this leads to an improvement of the ELEvoHI prediction accuracy. 

\section{Discussion}
\label{sec:discussion}

Data assimilation using the current STEREO/HI beacon images---which are the only data available in near real time---is unfortunately almost infeasible. Due to the numerous data gaps, it is difficult to track the CME across the entire HI1 FOV (up to $24^\circ$). As this study shows, with the current ELEvoHI approach, HI-based predictions outperform those not using HI data only when at least $35^\circ$ of elongation are utilised.
However, we are working on improving these data with the help of machine learning methods. A first step of this approach is presented in \citeA{lel25}.
Despite the poor quality of STEREO's near real-time HI data we are optimistic that the PUNCH mission will enable HI data assimilation in real-time. 
The PUNCH spacecraft, launched in March 2025, aims to significantly enhance CME forecasting by providing continuous 3D observations of the solar wind and corona. PUNCH's four small satellites capture wide-field images offering an additional vantage point. This mission is designed to provide a comprehensive view of how solar wind structures evolve, allowing for more accurate tracking of CMEs and their trajectories towards Earth.
Combining PUNCH data with existing observations from STEREO-A/HI could further improve forecasting accuracy by providing complementary views of CME propagation, helping to better estimate CME direction, speed, and shape. PUNCH's continuous monitoring and advanced imaging capabilities, e.g.\ the observation of different polarisation directions or its FOV covering $360^\circ$ around the Sun, could help overcome these limitations, offering a more reliable and detailed real-time data stream for operational space weather forecasting.

Improving the prediction accuracy of single, isolated CME-events is a first step towards improving space weather prediction itself. However, especially CME-CME interaction events cause the most intense geomagnetic disturbances \cite<i.e.>{liu14,koe22}. With our enhanced observation capabilities in the near future, we might be able to concentrate especially on improving the modelling of interaction events.

\section{Conclusions}

In this study, we systematically test the influence of various factors on the accuracy of CME arrival time predictions with ELEvoHI. First, we investigate the impact of HI data assimilation on the performance of ELEvoHI. Our results indicate that incorporating HI observations at least as far out as $35^\circ$ elongation improves predictions compared to those not utilising HI data. However, real-time assimilation remains challenging due to the limitations of current STEREO/HI beacon data, which suffer from frequent data gaps. With improved data quality from future missions like Vigil, real-time HI data assimilation will become feasible and significantly enhance space weather forecasting.  

We also examine how different individuals tracking CMEs in HI images influence the results and find that especially for shorter track lengths the variations in manual tracking lead to remarkable differences between the resulting prediction accuracies, while for longer tracks the differences decrease. Standardised training or automated tracking methods are necessary to minimise subjectivity.  

Additionally, we explore the influence of different methods for deriving input parameters for ELEvoHI, comparing GCS- and FPF-based approaches. While both methods have their strengths and limitations, discrepancies between them can lead to significant differences in predicted arrival times, highlighting the importance of carefully selecting and validating input parameters. \citeA{har23} noted that several Earth-directed events observed by HI in 2011 were missing in coronagraph data. This is supported by our study, where only 9 out of 15 CMEs detected at L1 can be fitted in coronagraph observations (either due to data gaps or too faint structures), emphasising the need for high-quality HI data available in real time from outside the Sun--Earth line. A significant practical advantage of the HI-alone ELEvoHI/FPF technique is its extremely low observational demand---it relies solely on one instrument on one spacecraft. This stands in clear contrast to the GCS method, which requires two vantage points and complementary measurements from multiple instruments to identify CME source regions.

A key finding of our study is that deterministic runs, i.e.\ those ensemble members directly based on the results from GCS or FPF, rather than the ensemble median or mean often yield a better prediction accuracy. This suggests that, instead of relying on the ensemble median/mean, it might be beneficial to use the deterministic result as the central prediction and define asymmetric error boundaries around it based on the ensemble spread. This approach could provide a more realistic representation of uncertainty in CME predictions.    

Looking ahead, the transition to real-time space weather forecasting based on HI data will require substantial improvements in both observation and modelling capabilities. The PUNCH mission, with its continuous imaging of the heliosphere from Earth orbit, has the potential to add additional value to HI-based CME prediction. Additionally, advancements in solar wind modelling, particularly through real-time data assimilation techniques, will be crucial for refining CME propagation modelling.  

Finally, this study is paving the way for CME predictions based on heliospheric imager observations from the future Vigil mission at L5. Since the CMEs analysed here are all observed from a vantage point around L5, our findings provide valuable insights into the potential of Vigil to enhance space weather forecasting. By continuously monitoring CMEs from a stable position at L5, Vigil will offer a complementary perspective to Earth-based and L1 observations, improving our ability to predict CME arrival times, speeds, and impact probabilities in real-time.

%
%

\section{Open Research Section}
\label{sec:data_access}
The STEREO/HI data used was downloaded from \url{https://stereo-ssc.nascom.nasa.gov/pub/ins_data/secchi/L0/} and processed with the Python routines available under \url{https://doi.org/10.5281/zenodo.17817626}. The tool used to perform the tracking of the CME fronts in time-elongation maps can be retrieved via \url{https://doi.org/10.5281/zenodo.17864921}. HELCATS HICAT can be accessed under \url{https://www.helcats-fp7.eu/catalogues/wp2_cat.html}.
The ELEvoHI version used for this study is published under \url{https://doi.org/10.5281/zenodo.17865833}. All model runs, the comparisons of the different time-elongation tracks and the code to produce the figures are available at \url{https://doi.org/10.5281/zenodo.17865833}.

\acknowledgments
This research was funded in whole or in part by the Austrian Science Fund (FWF) [10.55776/P36093] and [10.55776/P34437]. For open access purposes, the author has applied a CC BY public copyright license to any author-accepted manuscript version arising from this submission.
Funded by the European Union (ERC, HELIO4CAST, 101042188). Views and opinions expressed are however those of the author(s) only and do not necessarily reflect those of the European Union or the European Research Council Executive Agency. Neither the European Union nor the granting authority can be held responsible for them.
We acknowledge the Community Coordinated Modeling Center (CCMC) at Goddard Space Flight Center for the use of the CCMC CME Scoreboard (\url{https://kauai.ccmc.gsfc.nasa.gov/CMEscoreboard/}).
DB and JD recognise the support of the UK Space Agency for funding STEREO/HI operations in the UK.

%
\bibliography{tam_bib} 
%


%
%
%
%
%

\end{document}